\documentstyle [12pt] {article}

\begin{document}
\hfill {IPM-93-012}
\title {\large
{\bf SOME CORRELATORS OF $SU(3)_3$ WZW MODELS \\ ON HIGHER-GENUS RIEMANN
SURFACES} \author {Masoud Alimohammadi}}
\date{May 1993}
\maketitle
\begin{center}
Institute for Studies in
Theoretical Physics and Mathematics\\
P.O.Box: 19395-1795,
Tehran, Iran\\
Email: ALIMOHMD @ IREARN. bitnet
\end{center}

\vspace {10 mm}
\begin {abstract}
{Using the conformal embedding on the torus, we can express some characters
of $SU(3)_3$ in terms of $SO(8)_1$ characters. Then with the help
of crossing symmetry, modular transformation and factorization properties of
Green functions, we will calculate a class of correlators of $SU(3)_3$
on arbitrary Riemann surfaces. This method can apply to all $k>1$ WZW models
which can be conformally embedded in some $k=1$ WZW models.}\\ \\
\end{abstract}
\noindent
{\large \bf I. Introduction}\\

Understanding of conformal field thoery, with infinite dimensional conformal
symmetry, and Wess-Zumino-Witten (WZW) models, with extra affine Kac-Moody
symmetry$^{[1-5]}$, demands treating them on arbitrary world sheet. Therefore
studying
WZW models on higher-genus Riemann surfaces (HGRS's) is one of the important
branches of the string theory.

So far the results that have been achieved toward the solution of the above
problem, can be summarized as follows (for brevity in the following we denote
the Green function and partition function by G.f. and p.f. respectively). Most
of the calculations have been done for the level-one WZW models, since these
are rather easier to handle. The equivalence of the level-one simply-laced
group with compactified free bosons, was used to calculate the G.fs. of the
primary field of these models on HGRS's $^{[6-8]}$. As an another approach, the
behavior of the correlators under zero-homology pinching (ZHP) and non-zero
homology pinching (NZHP) (or its factorization properties) can be used to
calculate the G.fs. By using this method, the $n$-point functions of the
primary
fields, and also a class of the descendant fields, was calculated for the
level-one simply-laced and also non-simply-laced $(SO(2N+1)_1)$ WZW models on
HGRS's $^{[9-11]}$.

Similar calculations for the higher-level $(k>1)$ WZW models, are more
important and of course more complicated.The only method that have been used
in these cases
is based on the Coulomb gas representation of these models.
The complete structure of these representations (that is their BRST cohomology
structure) was only found for $SU(2)_k$ $^{[12,13]}$, and therefore the
calculation of G.fs. was only done for this case (only the partition function
on genus-two was explicitly calculated $^{[14]}$).

As a result, searching for other methods of calculations of $k>1$ WZW models
is of great importance. In this paper, we will present another approach which
can be applied to a large class of $k>1$ models. This method is based on:
conformal embedding, crossing symmetry, modular transformation and
factorization
properties of G.fs.

Conformal embedding allows one to express some characters of a class of $k>1$
models $(H_k)$ in terms of the characters of some specific level-one WZW models
$(G_1)$. By $H_k$ we mean a WZW model with group H and level k. It is of
crucial
importance to note that this equivalence of p.fs. of $H_k$ and $G_1$ is
restricted to genus-one (torus). By increasing the genus, the equivalence is
lost due to the fact that the difference between the number of diagrams,
in the two theories, increases rapidly. Therefore we allow to use the conformal
embedding only in the first step of our calculations (that is on the torus),
and to find the higher-genus correlators we must seek another techniques
(those mentioned above).

Here we will study a simple example of this kind, that is $SU(3)_3$ (which
can be conformally embedded in $SO(8)_1$), but our method can be easily
generalized to other cases which can be embedded in some level-one models.
In this papers, we will compute a class of G.fs. on arbitrary Riemann
surfaces.
This correlators correspond to the diagrams in which $\Phi_8,\Phi_{10}$ and
$\Phi_{\overline {10}}$ are external legs, $\Phi_8$ is the loop field and
$\Phi_1, \Phi_{10}$ and $\Phi_{\overline {10}}$ are propagators (the indices
denote the representations of $SU(3)$).

In sec. II we will briefly explain the conformal embedding and the branching
rules of $SU(3)_3$ in $SO(8)_1$. Sec. III is devoted to the calculation
of the fusion rules of $SU(3)_3$ and specifying the propagators and Sec.IV
to the calculation of genus-two p.fs. All the above mentioned G.fs. will be
calculated in sec.V. The necessary mathematical tools will be provided in the
appendix.\\ \\
{\large \bf II. Some $SU(3)_3$ Characters on the Torus}\\

Consider a subalgebra $h$ of a finite dimensional Lie algebra $g$. One can
associate to $h$, a Kac-Moody subalgebra $\hat {h}$ of an untwisted kac-Moody
algebra $\hat {g}$ by identifying the derivations $^{[15]}$ ($i.e.$ identifying
the basic gradations). Doing this, the central charges of the Kac-Moody
algebras get identified with a relative factor $j$ given by the Dynkin index
of the embedding of $h$ in $g$. This implies that every level $k$ highest
weight
$g$-module $L(\land )$ is reducible into a sum of level $\overline {k} =jk$
highest weight $h$-module $L(\overline {\land})$. In general this branching is
not finite,$i.e.$ the highest weight $g$-module $L(\land)$ is not finitely
reducible under $\hat {h}$. One can prove however that $L(\land)$ is finitely
reducible under $\hat {h}$ if and only if the corresponding Virasoro algebras
have equal central charge $^{[16]}$
\begin{equation}
c_g(k)=c_h(jk).
\end{equation}
In particular, it can be shown that, Eq.(1) can only be satisfied if $k$=1.
In this case $h_j$ is conformally embedded in $g_1$. All of these embeddings
has been classified $^{[17]}$.

Since $SU(N)$ is a subgroup of $SO(N^2-1)$ and the Dynkin index of this
embedding is N, $SU(N)_N$ can be embedded (conformally) in $SO(N^2-1)_1$.
As the first nontrivial example, we will consider the N=3 case.

As mentioned above, the characters of $SO(8)_1$ can be decomposed in terms of
$SU(3)_3$ characters. But before discussing the branching rules, we must
specify
the primary fields of these two models. As is well known, the primary fields of
$SU(N)_k$ are those representations for which the width of the corresponding
Young tableaux are less than or equal to $k$. Therefore the primary fields of
$SU(3)_3$ are:
\begin{equation}
\Phi_1,\Phi_3,\Phi_{\overline {3}},\Phi_6,\Phi_{\overline
{6}},\Phi_8,\Phi_{10},
\Phi_{\overline {10}},\Phi_{15},\Phi_{\overline {15}} \ \ .
\end{equation}
The indices show the dimension of representation. The conformal weights of
primary fields are
\begin{equation}
h_r=\frac {C_r/\psi^2}{2k/\psi^2+g} \ \ ,
\end{equation}
where $C_r$ is the quadratic Casimir of representation, $\psi$ is the highest
root, $k$ is the level and $g$ is the dual coaxter number. For
representation $(m,n)$
of $SU(3),C_{mn}/ \psi^2=\frac {1}{3}(m^2+n^2+mn+3m+3n)$ and $g$=3. By choosing
$\psi^2=2$, we find the following conformal weights for the primary fields
of $SU(3)_3$
\begin{equation}
h_1=0,h_3=h_{\overline {3}}=\frac {2}{9},h_8=\frac {1}{2},h_6=h_{\overline {6}}
=\frac {5}{9},h_{10}=h_{\overline {10}}=1,h_{15}=h_{\overline {15}}=\frac
{8}{9} \ \ .\end{equation}
The primary fields of $SO(8)_1$ can also be determined as
$\Phi_0,\Phi_v,\Phi_s$
and $\Phi_{\overline {s}}$ where the indices stand for 0=trivial, $v$=vector
and
$s(\overline {s})$=spinor (antispinor) representations of $SO(8)$. The
conformal
weights of these primary fields are: (see Eq.(3))
\begin{equation}
h_o=o,h_v=h_s=h_{\overline {s}}=\frac {1}{2} \ \ .
\end{equation}
Comparison of Eqs.(4) and (5), shows that if we expand the character $\chi_0$
in terms
of the characters of $SU(3)_3$, only those characters enter which the
corresponding
primary fields have integer conformal weights. To see this let's remind the
defi
nition
of the character of a primary field of conformal dimension $h$.
\begin{equation}
\chi_h(q)=q^{-c/24}trq^{Lo}=q^{-c/24}(a_1q^h+a_2q^{h+1}+\cdots ) \ \
\ , \ \ q=e^{2\pi i \tau} \end{equation}
Neglecting the factor $q^{-c/24}$ (as is the same for both models, i.e.
$c_{SO(8)_1}=c_{SU(3)_3}=4)$, the powers of q in $\chi_0(q)$ are all integers
(as $h_0=0$) and therefore only the primary fields of integer conformal weight
can take part in $\chi_0$'s decomposition. Direct calculation shows that
$^{[18]}$ \begin{equation}
\chi_0=\chi_1+\chi_{10}+\chi_{\overline {10}} \ \ ,
\end{equation}
and in the same way
\begin{equation}
\chi_s=\chi_{\overline s}=\chi_8 \ \ .
\end{equation}
Using these relations, it is possible to write the above characters of
$SU(3)_3$
 in terms of
$SO(8)_1$ characters.\\
The characters of $SO(8)_1$ can be calculated in two ways, either by using the
s
tring
functions of $SO(8)_1$, in the same way that the characters of $SO(2N+1)_1$ was
found in Ref.[11], or by using the free field representation of this theory
$^{[
19]}$. By
either way the result is
\begin{equation}
\chi_0(\tau)=\frac {1}{2\eta^4(\tau)} \left\{ \Theta^4\left[ \begin{array}{l}
0 \\ 0
\end{array} \right] (0 \vert \tau)+\Theta^4\left[\begin{array}{l}
0 \\ 1/2
\end{array} \right] (0 \vert \tau)\right\}
\end{equation}
$$\hspace {3.2cm}\chi_{v}(\tau)=\frac
{1}{2\eta^4(\tau)}\left\{\Theta^4\left[\
\begin{array}{l} 0 \\ 0
\end{array} \right] (0 \vert \tau)-\Theta^4\left[\begin{array}{l}
0 \\ 1/2
\end{array} \right] (0 \vert
\tau)\right\} \hspace {25mm} (9-b) $$
$$\hspace {4.2cm} \chi_s(\tau)=\chi_{\overline {s}}
(\tau)=\frac {1}{2\eta^4(\tau)} \Theta^4\left[\begin{array}{l}
1/2 \\ 0
\end{array} \right] (0 \vert \tau) \ .\hspace {3.8cm} (9-c)  $$
 Since $ \Theta ^4 \left[\begin{array}{l}
1/2 \\ 0
\end{array} \right] (0 \vert \tau)=\Theta^4\left[\begin{array}{l}
0 \\ 0
\end{array} \right] (0 \vert \tau)-\Theta^4\left[\begin{array}{l}
0 \\ 1/2
\end{array} \right] (0 \vert \tau),$ we have:
\begin{equation}
\chi _v = \chi _s= \chi _{\overline {s}}= \frac {1}{2 \eta ^4(\tau)} \Theta
^4 \left[ \begin{array}{l}
1/2 \\ 0
\end{array}\right] (0 \vert \tau).
\end{equation}
On the other hand, the bosonic p.f. on the torus is
\begin{equation}
Z_B(\tau)=(Im \tau)^{-1/2} | \eta (\tau)|^{-2},
\end{equation}
therefore
\begin{equation}
\hat {Z} _8(\tau) = \frac {(Im\tau)^2}{2}\biggl\vert \Theta \left[
\begin{array}
{l}
1/2\\ 0
\end{array}\right] (0 \vert \tau)\biggr\vert ^8
\end{equation}
$$\hspace {25mm}(\hat {Z}_1+ \hat {Z}_{10}+ \hat
{Z}_{\overline {10}})(\tau)= \frac {(Im
\tau)^2}{2} \biggl\vert \Theta ^4 \left[ \begin{array}{l}
0\\ 0
\end{array}\right] (0 \vert \tau)+ \Theta ^4 \left[ \begin{array}{l}
0\\ 1/2
\end{array}\right] (0 \vert
\tau)\biggr\vert ^2,\hspace {10mm}(12-b) $$
where by $\hat {Z}$ we mean
\begin{equation}
\hat {Z} = \frac {Z}{Z^c_B}
\end{equation}
The reason of division of p.f. by $Z_B^c$ is to write the p.f. in a metric
independent way $^{[20]}$,
and in fact only this ratio is well defined as the p.f. and must be
generalized to HGRS's.

Before concluding this section, it is worth noting that the decompositions
simil
ar to Eqs. (7) and (8)
also occur in other cases, and the techniques which we will develop in this
pape
r, can be extended to apply to these
more general cases also. In the following we quote some of these branching
rules
 $^{[18]}$:
\begin{equation}
\begin{array}{lll}
ch_{10} & = ch_{\overline {10}} & = \chi _{10}\\
ch_{5} & = ch_{\overline {5}} & = \chi _{5},
\end{array}
\end{equation}
where $ch_i \in SU(5)_1$ and $\chi _i \in SO (5)_2$,
\begin{equation}
ch_{16}=ch_{\overline {16}}=\chi_{16}
\end{equation}
with $ch_i \in SO(10)_1$ and $\chi_i \in SO(5)_3,$ etc.\\ \\
{\large \bf III. $SU(3)_3$ Fusion Rules}\\

In this section, we will find $SU(3)_3$ fusion rules of the conjugate
fields, by orthogonal polynomials technique, developed by Gepner $^{[21]}$.
Let us briefly review this technique.

Let $\mu_i(i=1,\cdots ,N)$ be the fundamental weights of $SU(N)$. Then any
arbitrary representation with highest weight $\mu$, can decompose as $\mu =
{\displaystyle \sum_{N-1}^{i=1}}a_i\mu_i$. In this way, we can denote any
primary field of $SU(N)_k$ by $[a_1,\cdots ,a_k]$ where $0\leq a_1 \leq a_2
... a_k \leq N-1$. Now if we denote the fully antisymmetric
(fundamental) reprsentations of $SU(N)$ by
\begin{equation}
{\overline c}_r=[0,0,\cdots ,0,r]\ \ \ \ \ \ \ r=0,1,\cdots , N-1,
\end{equation}
whose Young tableaux is\ \ \ \ \ \ \ \ , then Gepner has shown that there
exist a natural map from $[a_1,\cdots , a_k]$ to the polynomials of $N$
indeterminates $\bar c_i$, such that one can express this representation as
follows: \begin{equation}
[a_1,\cdots , a_k]= det_{\scriptstyle 1 \leq i,j \leq k \atop \scriptstyle \ \
} {\overline c}_{a_i+i-j}
\end{equation}
The $det$ stands for the deteminant of the matix $A_{ij}={\overline
c}_{a_i+i-j}$,
which is defined by the convention ${\overline c}_0={\overline c}_N=1$ and
${\overline c}_i=0$ for $i>N$ or $i<0$. It can also be proved that the fusion
rule of ${\overline c}_r$ with other representations can be read from the
following equation$^{[21]}$:
\begin{equation}
\bar c_r.[a_1,...,a_k]=\sum_{\scriptstyle a_i \leq b_i \leq a_{i+1} \atop
\scriptstyle \sum b_i=r+\sum a_i } [b_1,...,b_k]
\end{equation}
In this equation $b_i$'s must be written mod$N$, and must be ordered as
$0\leq b_1\leq b_2\cdots \leq b_k$. In the RHS of Eq.(18) each representation
must be considered only once.

Now let us apply this method to $SU(3)_3$. In this case, the variables of
polynomials are two fundamental representation of $SU(3)_3$, that is
$3=[0,0,1]=x$ and ${\overline 3}=[0,0,2]=y$. Then Eq.(17) allows us to write
for each representation of $SU(3)_3$ a corresponding polynomial\\
\begin{equation}
\begin{array}{l}
1=[0,0,0]=0\ \ \ ,\ 3=[0,0,1]=x\ \ \ ,\ {\overline 3}=[0,0,2]=y\ \ \ ,\ 6=[
0,1,1]=x^2-y\\
{\overline 6}=[0,2,2]=y^2-x\ \ ,\ 8=[0,1,2]=xy-1\ ,\ 10=[1,1,1]=x^3-2xy+1\\
{\overline {10}}=[2,2,2]=y^3-2xy+1\ \ ,\ 15=[1,1,2]=x^2y-x-y^2\ \ ,\ {\overline
{15}}=[1,2,2]=xy^2-x^2-y
\end{array}
\end{equation}
Now it is possible to find the fusion rules by using Eq.(18). For example:
$$\Phi_6\times\Phi_{\overline 6}=(x^2-y).[0,2,2]$$
but
$$y.[0,2,2]=[0,1,2]+[2,2,2]$$
$$x.[0,2,2]=[0,0,2]+[1,2,2]$$
$$x^2.[0,2,2]=x.[0,0,2]+x.[1,2,2]=[0,0,0]+2[0,1,2]+[2,2,2]$$
therefore
$$\Phi_6\times\Phi_{\overline 6}=[0,0,0]+[0,1,2]=\Phi_1+\Phi_8$$
In this way we obtain the following fusion rules of the conjugate
representations of $SU(3)_3$
\begin{equation}
\begin{array}{l}
\Phi_3 \times \Phi_{\overline 3}=\Phi_1+\Phi_8\ ,\ \Phi_6\times\Phi_{\overline
6}=\Phi_1+ \Phi_8\ ,\ \Phi_{10} \times \Phi_{\overline {10}}=\Phi_1\\
\Phi_8\times\Phi_8=\Phi_1+2\Phi_8+\Phi_{10}+\Phi_{\overline {10}}\ ,\ \Phi_{15}
\times\Phi_{\overline {15}}=\Phi_1+\Phi_8
\end{array}
\end{equation}
These fusion rules are sufficient to know all the multiloop diagrams of
$SU(3)_3
$.
The reason is the following: As is obvious from sewing procedure, we can
construct
any loop by sewing (fusing) two legs of a three-point vertex, only when their
corresponding representations are conjugate. Therefore to know the allowed
propa
gators
of a theory, it is enough to only consider the fusion rules of the conjugate
fie
lds
(Fig.1). Eq.(20) shows that in $SU(3)_3$ diagrams on HGRS's, only the fields
$[\Phi_1],[\Phi_8],[\Phi_{10}]$ and $[\Phi_{\overline {10}}]$ are propagators.
By $[\Phi_i]$, we mean the family of $\Phi_i$, that is the primary field and
all
the descendant fields which can be built on $\Phi_i$.\\ \\
{\large \bf IV. Genus-two Partition Functions}\\

Everything is now ready for calculation of multiloop diagrams. As is  clear
from Eq.(20), in the genus-two p.fs., whenever $\Phi_8$ is the field which
circulate in the loops, only the fields $\Phi_1,\Phi_8,\Phi_{10}$ or $\Phi_
{\overline {10}}$ can propagate between the two loops (Fig.2). (From now on,
whenever we do not specify the fields of the loops, we mean that they are
$\Phi_8$).

For $\Phi_i=\Phi_1$ (Fig.3), the calculation of the p.f. is straightforward.
As was extensively discussed in Refs. [9-11], the ZHP and NZHP behaviors of
the G.fs. are very important for the calculation of the correlators of a
conformal field  theory on HGRS's. In particular, in the cases which the
identity operator is propagator, these limits determine the p.fs. uniquely.
Therefore we can find the p.f. of Fig.3 as:
\begin{equation}
\hat {Z}^1_{88}(\Omega )=\frac {(det\ Im\ \Omega)^2}{4}\ \biggl\vert \Theta
\left[
\begin{array}{ll}
1/2 & 1/2 \\
0 & 0
\end{array} \right] (01\Omega)\ \biggr\vert ^8,
\end{equation}
where by the index $i$ in $\hat {Z}^i_{88}$, we mean that the propagator of
the p.f. (in Fig.2) is $\Phi_i$. In Eq.(21) $\Omega$ is the period matrix
of the Riemann surface and $\Theta \left[\begin{array}{ll}
a_1 & a_2 \\
\beta^1 & \beta^2
\end{array} \right]$
are the genus-two Riemann theta functions (Eq. A.5). It can be shown that Eq.
(21) behaves correctly under ZHP and NZHP limits, that is, it leads in these
limits to:
\begin{equation} \begin{array}{l}
\hat {Z}^1_{88} (\Omega)\ \stackrel {ZHP}\longrightarrow \hat
{Z}_8(\tau_1)\hat {Z}_8(\tau_2)+O(t)\\ \\
Z^1_{88}(\Omega )\ \stackrel {NZHP}\longrightarrow <\Phi_8(x_1)\Phi_8(x_2)>
(\tau) |t|^{1/2+1/2}+\cdots
\end{array} \end{equation}
In the above relations, we have used Eqs. (A.6),(A.7) and (12a). As $h_8=
\frac {1}{2}$, it is obvious that Eq.(21) behaves as predicted by Eqs. (A.2)
and (A.3).

In this way we could find one of the p.fs. of Fig.(2). Now what about the
others. The main reason that we could write Eq.(21), was its easy ZHP
behaviour (Eq. A.2), and this came from the fact that in the corresponding
diagram, the propagator is the scalar field $(\Phi_1)$. But this is not
true for other diagrams and we must look for other techniques to calculate
them.\\
 The method that we are going to apply it to the remaining diagrams,
is based on the using the modular transformation properties and crossing
symmetry of Green functions. So lets begin with the definition of the
generators of the modular transformations.
As is known, the Dehn twist about the cycle $c$, $D_c$, is equivalent to
cutting the surface along the cycle $c$, rotating one cut end by $2\pi$, and
then glueing the surface back together again. Now if we choose a homology
basis $(a_i,b_i)$, then it can be shown that $^{[22]}$ the generators of
modular transformation are in a 1-1 correspondence with the set of Dehn twists
about the cycles $a_i,\ a^{-1}_ia_{i+1}$ and $\prod^g_{i=1}a^{-1}_i b_i
a^{-1}_i.$

Now consider the genus-two diagram of Fig.4. Cutting this surface along
the $a$ cycles, hence transforming it to a sphere with four puncture, and
using the crossing symmetry (which interchanges the $t$ and $s$ channels) we
arrive at Fig.5. The sum over $\ell$ is restricted by fusion rule. Next
consider the cycle $a^{-1}_1 a_2$ of Fig.4 (which is the same as the
cycle $c$ in Fig.5) and lets apply the Dehn twist $D_c$ in both sides of
Fig.5. There is a crucial difference between these two twists. In the
LHS, there is no specific field which propagates along the cycle $c$, and
therefore $D_c$ is nontrivial. But in the RHS, there is such a field,
$\Phi_{\ell}$, and therefore $D_c$ is trivial and only produce the phase
factor $e^{2i \pi h_\ell}$.
In the following, we will use the above mentioned points to find the
unknown diagrams.

Consider the diagram of Fig.3. If we cut the loops, we find a four-point
function which can be expanded as follows: \\ \\ \\
\begin{equation}
\end{equation} \\ \\
The propagators in the RHS is determined by fusion rule. Note that at this
stage, and before any twisting, it is clear that $c_1 =1 $ and $c_8=c_{10}
= c_{\overline {10}} =0 $. If we again glue the external legs of Eq.(23), we
find the following relation between the p.fs.
\begin{equation}
\hat {Z} _{88} ^1 ( \Omega ) =
\sum_{\scriptstyle \ell =1,8 \atop \scriptstyle
10 , \overline {10}}
c_{\ell } (\Omega ) \hat {Z} _{88}^{\ell } (\Omega )
\end{equation}
Now lets twisting both sides about the cycle $c$ $(D_c)$. Any modular
transformation
is specified by four $g \times g$ matrices $A,B,C$ and $D$ which determine the
transformation of the period matrix
\begin{equation}
\Omega \rightarrow \stackrel {\sim }{ \Omega } = (A \Omega + B) (C \Omega
+D)^{-
1}
\end{equation}
In the case of $D_{a_1 ^{-1} a_2 }$ these matrices are (in $g=2$):
$$A=D= \left ( \begin{array}{ll}
1 & 0 \\
0 & 1
\end {array} \right ), B = \left ( \begin{array}{ll}
-1 & 1 \\
1 & -1
\end {array} \right ), C = 0.$$
By Using the above matrices, and Eq.(A.8), we find that $\hat {Z} _{88}^1
(\Omega )$ is transformed as (by Eq.(21)):
\begin{equation}
\hat {Z}_{88}^1 (\Omega ) \stackrel {D_{a_1^{-1}a_2}} {\longrightarrow }
\frac {(det \ Im \Omega )^2}{4} \biggl\vert \Theta \left[ \begin {array}{ll}
1/2 & 1/2\\
1/2 & 1/2
\end{array} \right] (0 \vert \Omega ) \biggr\vert ^8
\end{equation}
The transormation of the RHS of Eq.(24) will only leads to an
$e^{2 \pi ih _\ell }$ phase factor, and therefore Eq.(24) leads to
$$\frac {(det \ Im \Omega )^2}{4} \biggl\vert \Theta \left[ \begin {array}{ll}
1/2 & 1/2\\
1/2 & 1/2
\end{array} \right] (0\vert \Omega ) \biggr\vert ^8= \sum _{\ell } c' _\ell e
^{2 \pi i h_\ell } \hat {Z}_{88}^{\ell } (\Omega )$$
\begin{equation}
=c_1 '\hat {Z}_{88}^1-c'_8 \hat {Z} _{88}^8 + c'_{10} \hat {Z}_{88}^{10} + c'
_{\overline {10}} \hat {Z} _{88} ^{\overline {10}} \ \ . \end{equation}

To determine the coefficients $c'_\ell$, we study the ZHP behavior of
both sides of Eq.(27). Note that both sides must behave identically. From
Eq.(A.6) one can easily verify that in the ZHP limit, the pinching parameter
$t$ appears in the LHS of Eq.(27) with the power $t^8$. On the other
hand, in the ZHP limit of a p.f. of a primary field with conformal
weight $h,t$ appears with the power $(2h+n)$,
$(n=0,1,2, \cdots )$, where $n$ denotes the level of its descendants (we have
su
pposed the
left-right symmetry for p.f. $i.e. \ \ h=\overline {h}$). Therefore one finds
that Eq.(27) can be written in this limit as:
\begin{equation}
t ^8 \sim c'_\circ (t ^\circ +t^2 + \cdots ) + c' _8 (t^1+t^3 + \cdots ) +
c'_{10} ( t^2 + t^4 + \cdots ) + c'_{\overline {10}} (t^2 + t^4 +
\cdots ) \end{equation}
where we have suppressed the coefficient of $t^n $'s for simplicity. Using the
fact that the ZHP limit of $\hat {Z} _{88} ^1$ appears as $t^\circ +t^2 +
\cdots
$ (Eq.22), the
above relation leads:
\begin{equation}
c'_\circ = c' _8=0
\end{equation}
To determine $c' _{10}$ and $c'_{\overline {10}}$, we use the fact that the
representations $10$ and $\overline {10}$
appears symmetricaly in Eqs.(4,12.b and 20) and therefore $c'_{10} =
c'_{\overline {10}}$ and
$\hat {Z}_{88}^{10}= \hat {Z} _{88}^{\overline {10}}$. Without loss of
generality we set $c'_{10} = c'_{\overline {10}}=1/2$ and finally:
\begin{equation}
\hat {Z}_{88}^{10}= \hat {Z} _{88}^{\overline {10}} = \frac {(det \ Im \Omega
)^2}{4} \biggl\vert \Theta \left[ \begin {array}{ll} 1/2 & 1/2\\
1/2 & 1/2
\end{array} \right] (0 \vert \Omega ) \biggr\vert ^8
\end{equation}

Let us  convince ourselves, why in the ZHP limit, the leading terms of
$\hat {Z}_{88}^{10}$ and $\hat {Z} _{88}^{\overline {10}}$ is $t^8$. This
means,
 with the help of
Eq.(A.1), that the first nonvanishing one-point function of $< [ \Phi _{10}]
>$,
 On the
torus, appears at level three. From the general properties of the Kac-Moody
alge
bras, one can show the
following condition for the one-point functions of the descendant
fields$^{[10]}$ :
\begin{equation}
<0|J_{-m_1}^{\beta _1} \cdots J_{-m _k}^{\beta _k} \Phi_\Lambda ^{\lambda } |
0>=0
\ \ \ \ {\rm unless } \ \ \ \  \lambda + \sum _{i=1}^{k} \beta _i =0,
\end{equation}
where $\lambda $ is a specific weight of a representation $\Lambda$. This
relation can help us to interpret our results.

In our case $\Lambda =10$. Suppose that $\lambda $ is the highest weight
of this
representation, that is $\lambda =3 \nu _1 (\nu _1$ is the highest weight of
representation $3,i.e.,R_3$). Therefore
the selection rule (31) implies that the G.fs. is zero, unless
$\sum ^k_{i=1} \beta _i = -3 \nu _1$. One of the solution of this relation
is $\beta _1=\beta _2 = -\alpha _1$ and $ \beta _2 = - \alpha_2$, where
$\alpha _i$'s are the simple roots of $SU(3)$. Hence we encounter the
descendant fields of the form $J^{-\alpha  _1}_{-m_1} J^{-\alpha _1}_{-m_2}
J^{-\alpha _2}_{-m _3} \Phi _{10}^{3\nu}$. These fields have conformal
weights $1+m_1+m_2+m_3$. As the minmum value of $m_i$'s are one,
so the minimum level of the non-zero one-point function of the descendant
fields of $\Phi _{10}$ is three. This is exactly the same result that we found
in Eq.(18), that is the first non-zero one-point function of $\Phi _{10}$ (with
minimum conformal weight) appear at level three.

There is yet a remaining question. In our analysis, we only consider a special
set of solutions, that is $\lambda = 3 \nu$, and a special decomposition of
$\sum \beta _i$
(in terms of the roots). How we can use this, to explain the general result of
Eq.(28),
which does not depend on any specific weight and root of descendant fields. The
answer is: There is some indications that $^{[10]}$ the G.f. of descendant
$J_{-m_1}^\beta  \cdots J_{-m_k}^{\beta _k } \Phi^\lambda_\Lambda $
are equal, as long as $\lambda + \sum ^ k_{i=1} \beta _i$ and $\sum ^{k}_{i=1}
m
_i$
are the same. So our result about the necessity of level three, is not
restricte
d
to the above case which we considered, and is correct for the general case.

Let us conclude this section with some comments about the information that are
contained in the second part of Eq.(12), that is (12.b). Unfortunately we can
not separate out the $\hat  {Z} _1 , \hat {Z} _{10}$ and
$\hat {Z}_{\overline {10}}$ from
this equation and therefore it only helps us to find a combination of p.fs. On
H
GRS's.
For instance, consider the genus-two p.f. It is obvious from Eq.(20) that in
the
diagrams in which the fields $\Phi_1,\Phi_{10}$ and $\Phi_{\overline {10}}$ are
their loop fields,
only the field $\Phi _1$ can propagate between successive loops. This makes the
calculation of these diagrams doable. Using a similar reasoning which led to
Eq.(21), we can find the following sum of p.fs.:
$$\sum _{i,j=1,10,\overline {10}} \hat {Z}^1_{ij} (\Omega ) = \frac {(det \ Im
\ \Omega )^2}{4}$$ \begin{equation}
\left | \left \{ \Theta ^4 \left[ \begin{array} {ll}
0 & 0 \\
0 & 0
\end{array} \right] + \Theta ^4 \left[\begin{array} {ll}
0 & 0 \\
0 & 1/2
\end{array} \right] + \Theta ^4 \left[\begin{array} {ll}
0 & 0 \\
1/2 & 0
\end{array} \right] + \Theta ^4 \left[\begin{array} {ll}
0 & 0 \\
1/2 & 1/2
\end{array} \right] \right \} (0 \vert \Omega ) \right | ^2
\end{equation}
It can be easily checked that the above relation will produce the correct ZHP
behavior, that is
$$\sum _{i,j=1,10,\overline {10}} \hat {Z} ^1 _{ij} (\Omega )
\stackrel {ZHP} {\longrightarrow} \sum _{i=1,10,\overline {10}} \hat {Z} _i
(\tau_1) \sum _{j=1,10,\overline {10}} \hat {Z} _i (\tau_2)$$
The generalization of this combination to HGRS's, that is
$\sum _{i_1,\cdots , i_g=1,10,\overline {10}} \hat {Z} ^{11 \cdots 1} _{i_1
\cdots i_g} (\Omega_g )$, can be found in a similar way.\\ \\
{\large \bf V. Higher-Genus Correlators}\\

By similar techniques which led to Eq.(30), one can compute the higher-genus
p.fs. We
first consider the genus-three, in detail, and then give the results for
aritrar
y
genus p.fs. Here after, we will only consider the diagrams with $\Phi _1, \Phi
_
{10}$ and
$\Phi _{\overline {10}}$ as propagator (and also $\Phi _8$ as the loops
fields)

In genus three there are nine diagrams of this type (Fig.6). In five of them,
there is, at least, one $\Phi _1$ propagator and the resulting p.fs. can be
writ
ten by
considering its ZHP behavior. For example
\begin{equation}
\hat {Z}^{10,1}_{888} = \frac {(det \ Im \ \Omega )^2}{2^3}
\left | \Theta \left[ \begin{array} {lll}
1/2 & 1/2 & 1/2\\
1/2 & 1/2 & 0
\end{array} \right] (0 \vert \Omega) \right |^8
\end{equation}

In the remaining four diagrams both of the propagators are $10$ or $\overline
{1
0}$,
and these must be calculated in the same way which led to Eq.(30). But thanks
to
the equality of the contributions of $10$ and $\overline {10}$, all these four
diagrams reduces to $\hat {Z} ^{10,10}_{888}$ and only this p.f. must be
calculated.

Cosider $\hat {Z} ^{10,1}_{888}$
and apply the previous procedure to one of its loops:\\ \\ \\ \\ \\ \\ \\ \\ \\
\\
or:\\ \\
\begin{equation}
\end{equation}
\\  Obviously before any Dehn twisting, we have $c_1=1$ and $c_i
=0 (i \not= 1)$.
Now lets Dehn twist about the cycle $a_2^{-1} a_3$. The LHS of (34) changes as
\begin{equation}
\Theta \left[ \begin{array}{lll}
1/2 & 1/2 & 1/2 \\
1/2 & 1/2 & 0
\end{array} \right ]
\stackrel {D_{\alpha_2 ^{-1} \alpha _3}} {\longrightarrow} \Theta \left[
\begin{array}{lll}
1/2 & 1/2 & 1/2 \\
1/2 & 0 & 1/2
\end{array} \right ],
\end{equation}
and the RHS changes (trivially)
$$\sum c'_\ell e ^{2\pi ih _\ell} \hspace {40mm}$$
By considering the ZHP behavior of RHS of (33), and noting that the
pinching parameter appear with the power $4+4$, one can prove that
$c_1'=c'_8 =0$, and as $c'_{10} =c'_{\overline {10}}$, one finds
\begin{equation}
\hat {Z}^{10,10}_{888} (\Omega )= \frac {(det \ Im \ \Omega )^2}{2^3}
\left | \Theta \left[\begin{array} {lll}
1/2 & 1/2 & 1/2\\
1/2 & 0 & 1/2
\end{array} \right](0 \vert \Omega) \right |^8
\end{equation}
This method can be similarly applied to the general higher-genus cases and the
final result for the genus-$g$ partition function of Fig.7 is:
\begin{equation}
\hat {Z} ^{\bf m}(\Omega _g)=A_g \left | \Theta
\left[ \begin{array} {c}{\vec \alpha} \\ {\vec \beta }
\end{array} \right] (0 \vert \Omega _g)\right |^8
\end{equation}
where $A_g= \frac {(det \ Im \ \Omega _g )^2}{2^g}$ and ${\vec \alpha,
\vec \beta }$ and $ {\bf m}$ are $g,g$ and $g+1$ component objects
\begin{equation}
\begin{array}{ll}
\alpha _k = \frac {1}{2} & k=1,2,\cdots , g \\
\beta _k = \frac {1}{2} (1-\delta _{m_{k-1},m_k}) & k=1,2,\cdots , g \\
{\rm and} \ \ \ {\bf m}=(1,m_1,\cdots , m_{g-1}, 1). &
\end{array}
\end{equation}
Note that $m_i$ specifies the propagators (as shown in Fig.7) and takes the
value $1,10$ and $\overline {10}$. Note also that $\delta_{10, \overline
{10}}$ is to be set equal to one (due to the symmetry of these
representations).

To find the $n$-point functions, it is necessary to study the NZHP limits of
Eq.(37).
Various factors of this equation behave under NZHP of $g$-th loop as following
\begin{equation}
Z_B(\Omega _g) \stackrel {NZHP} {\rightarrow } Z_B(\Omega_{g-1})
(Im \ lnt)^{-1/2} |t| ^{-1/12}
\end{equation}
$$\hspace {42mm}det \ Im \Omega _g \stackrel {NZHP} {\rightarrow } (det \ Im
\Omega _{g-1}) (Im \ lnt)\hspace {35mm}(39-b) $$
$$\hspace {7mm}Z^{\bf m} (\Omega _g) \stackrel {NZHP} {\rightarrow }
Z_B^c (\Omega _{g-1}) |t|^{1/2 +1/2 - c/24 -c/24} {1\over 2} A_{g-1} \left |
\frac {2 \Theta \left[ \begin{array}{c}{\vec \alpha}  \\
{\vec \beta }
\end{array}\right] ( \frac {1}{2} \int _w ^z {\bf
u} |\Omega _{g-1})}{E^{1/4} (z,w)} \right | ^8 \hspace {5mm}(39-c) $$
Where in the last equation , (39-c) , use has been made of the first two one
and also Eq.(A.7). In
the above relations $c$ is centrd charge, ${\vec \alpha }$
and ${\vec \beta }$ have $g-1$ components and $E(z,w
)$ is
prime form on genus-$g$ Riemann surface $(\sum_g). {\bf u}
= (u_1 , \cdots , u_g)$
are differential one-forms on $\sum_g$. Eq.(39-c) shows that Eq.(37) behaves
correctly under NZHP. This is obvious by Eq. (A.3) and note that $h_8=
1/2$ (on cylinder
$h \rightarrow h-c/24)$. Therefore we will find the two-point function of $\Phi
_8 - \Phi _8$
on $\sum _g$ as (Fig.8):
\begin{equation}
<<\Phi _8 (z) \Phi _8 (w )>> _{\bf m}(\Omega_g)=
2^7A_g \left | \frac {\Theta \left[ \begin{array}{c}
{\vec \alpha}  \\ {\vec \beta }
\end{array}\right] ( \frac {1}{2} \int _w ^z {\bf
u} |\Omega _g)}{E^{1/4} (z,w)} \right | ^8 \end{equation}
\begin{equation}
<< \cdots >> = \frac { < \cdots > }{Z^c_B}
\end{equation}
and ${\bf m}= (1,m_1, \cdots , m_g)$. On can repeat this
NZHP technique for one the loops of Fig.8, and after obtaining the correct
power
for $t$, find the four-point function and again continue this procedure. In
this
way, one may find the $2n$-point function of $\Phi _8$ as following (Fig.9)
\begin{equation}
<< \prod ^n_{i=1} \Phi _8 (z_i)\Phi _8(w _i)>> _{\bf
 m} (\Omega _g)=2^{8-n} A_g \left | \frac { \xi _n \left[ \begin{array}{c}
{\vec \alpha}  \\{\vec \beta }
\end{array}\right]} {\prod ^n_{j=1} E^{1/4} (z_j, w _j)} \right
| ^8 \end{equation}
where ${\bf m } =(1,m_1 , \cdots, m_g)$ and
\begin{equation}
\xi _n \left[ \begin{array}{c}
{\vec \alpha} \\ {\vec \beta}
\end{array}\right] = \prod ^n_{p=2} (1+d_p) \prod ^n_{\scriptstyle i,j=1
\atop \scriptstyle i<j }
R_{ij}^{\gamma _i \gamma_j} \Theta \left[\begin{array}{c}
{\vec \alpha}  \\ {\vec \beta}
\end{array}\right] \left( \sum ^n_{i=1} \gamma _i \int _{w_i} ^{z_i}
{\bf u} |\Omega _g \right). \end{equation}
The index $i$ in $\gamma _i,z_i$ and $w _i$ refers to the $i$th external
leg (with arguments $z_i$ and $w _i$ in Fig.9), and all $\gamma _i$'s
are
$1/2$. $d_p$ is an operator which acts on any arbitrary function of $\gamma _p$
as following \begin{equation}
d_p f (\gamma _p)= exp [ \pi i (1- \delta_{m_{g+n-p},m_{g+n-p+1}})]
f(-\gamma _p), \end{equation}
and $R_{ij}$ is defined as follows:
\begin{equation}
R_{ij}= \frac {E(z_i,z_j)E (w _i , w _j)}{E(z_i, w _j) E (w _i z_j)}
\end{equation}

Finally we will describe the method of calculating the odd-point functions.
with the help of Eqs.(A.6) and (11), the ZHP limit of Eq.(30) will leads to:
\begin{equation}
Z_{88}^{10} \stackrel {ZHP} {\longrightarrow} 2^7 |\pi \eta ^2 (\tau_1)u_1
(p_1)|^8 \ \ |t|^{4+4} \ \  2^7 \ \ |\pi \eta ^2 (\tau_2) u_2 (p_2) |^8
\end{equation}
$u_1(u_2)$ is the holomorphic differential one-forms on the torus $T_1(T_2)$
(which is the result of pinching of $g=2$ surface) and $p_1(p_2)$ is the point
which is created on $T_1(T_2)$ by ZHP. By using Eqs.(46) and (A.1), One
can read the following one-point function of $\Phi _{10} (z)$ on torus (Fig.
10): \begin{equation}
<\Phi _{10} (z)> (\tau)=2^7 |\pi \eta ^2(\tau) u(z)|^8
\end{equation}
This is the one-point function of third-level descendant fields of $\Phi _{10}$
(as explained in the previous section).

Next consider the diagram of Fig.7 when $m_{g-1}=10$, and pinch the cycle
$a_g$. In this
way we will find the correct power for $|t|$ (that is $|t|^{4+4}$) and the
resul
ting relation
becomes the product of the $<\Phi _{10}> (\tau)$ (Eq.(47)) and $<\Phi _{10}>
(\Omega_{g-1})$. So we will find the $<\Phi_{10}> (\Omega _g)$ on HGRS's as
(Fig. 11) \begin{equation}
<<\Phi _{10} (z)>> _{\bf m}(\Omega _g)=Ag \left |\sum _{i=1}^g
u_i (z) \partial _j \Theta \left[\begin{array}{c}
{\vec \alpha}  \\
{\vec \beta }
\end{array}\right] (0 \vert \Omega _g) \right |^8
\end{equation}
where ${\bf m}=(1,m_1,\cdots , m_{g-1}, 10)$.

By the further ZHP and NZHP of the previous G.fs., one could
find all the Green function in which $\Phi _8$ is their loop fields, $\Phi _1,
\
Phi_{10}$
and $\Phi _{\overline {10}}$ are their propagators and $\Phi _8, \Phi_{10}$
or $\Phi _{\overline {10}}$ are external legs. \\ \\
{\bf Acknowledgement:} I would like to thank H. Arfaei and W. Nahm for fruitful
conversations. This work is partially supported by a grant from the sharif
Unive
rsity
of Technology.\\ \\
{\bf Appendix}

\

In this section, we will summarize some properties of the Green functions
and theta function on HGRS's which have been extensively used in the paper.
firs
t
we will review the factorization properties of correlators.

The factorization property of the genus-$g$ partition function was studied by
Fr
iedan
and Shenker$^{[23]}$.On a surface $\sum_{g,k}$ (a surface which becomes, on
removal of some node, two disconnected surfaces one of genus $k$ and one of
genus
$g-k$ , each with one puncture) in the limit of
$t \longrightarrow 0$ the partition function becomes:
$$\hspace {43mm}Z_g \buildrel ZHP \over \longrightarrow
<\phi(x_1)>_k<\phi(x_2)>_{g-k} \vert t \vert^{h+\bar h}\hspace {40mm}(A.1)$$
$<\phi(x)>_g$ is the normalized one-point function on $\sum_g$ , $\phi$ is the
field that propagates at that node and $t$ is the pinching parameter. In the
case that the propagator is $\phi_0$ , it reduces to:
$$\hspace {64mm}Z_g \buildrel ZHP \over \longrightarrow
Z_kZ_{g-k}\hspace {60mm}(A.2)$$
Under NZHP the partition function becomes $^{[21]}$:
$$\hspace {42mm}Z_g \buildrel NZHP \over \longrightarrow <\phi(x_1) \bar
\phi(x_2)>_{g-1} \vert t \vert^{h+\bar h}\hspace {45mm}(A.3)$$
$\phi$ is the field that propagates along the pinched handle.When this field
is $\phi_0$ Eq.(A.3) reduces to:
$$\hspace {65mm}Z_g \buildrel NZHP \over \longrightarrow Z_{g-1}
\hspace {57mm} (A.4)$$

Next we need to know the ZHP and NZHP behaviors of theta functions.The
Riemann theta function is defined as:
$$\hspace {15 mm}\Theta \left [ \begin{array}{c}  {\vec \alpha} \\ {\vec
\beta } \end{array} \right] (z \vert \Omega)
=\sum_{n \in Z^g}exp 2\pi i \left [ {1 \over 2} (n+{\vec \alpha})^t \Omega
(n+{\
vec \alpha})
+(n+{\vec \alpha})^t(z+{\vec \beta}) \right ] \hspace {10 mm}(A.5) $$
where ${\vec \alpha , \vec \beta}$ and $z$ are $g$-components objects, ${\vec
\a
lpha}$ and
${\vec \beta}$ belongs to $\{ 0,{1 \over 2} \}^g$ and \linebreak $z\in
C^g$.Unde
r ZHP
, $\sum_g$ divide to $\sum_1$ and $\sum_2$ (with $g=g_1+g_2$ ).Then$^{[24]}$:
$$\Theta \left[ \begin{array}{c}  {\vec \alpha} \\ {\vec \beta} \end{array}
\right] (\int^z_w {\bf v} \vert \Omega_g) \buildrel ZHP \over \longrightarrow
\Theta \left[ \begin{array}{c}
{\vec \alpha}_1 \\ {\vec \beta}_1  \end{array} \right]
(\int^z_w {\bf u} \vert \Omega_1)
\Theta \left[ \begin{array}{c}  {\vec \alpha}_2 \\ {\vec \beta}_2  \end{array}
\right] (0 \vert \Omega_2)$$
$$+t \omega_{z-w}(p_1) \left[ \sum^{g_1}_{i=1} u_i(p_1) \partial_i+
\sum^{g_1+g_2}_{j=g_1+1} u_j(p_2) \partial_j \right]
\Theta \left[ \begin{array}{c}  {\vec \alpha}_1 \\ {\vec \beta}_1 \end{array}
\right] (\int^z_w {\bf u} \vert \Omega_1)
\Theta \left[ \begin{array}{c}  {\vec \alpha}_2 \\ {\vec \beta}_2  \end{array}
\right] (0 \vert \Omega_2)$$
$$\hspace {3mm}+{1\over 2}t \left[ \sum^{g_1}_{i=1} u_i(p_1) \partial_i+
\sum^{g_1+g_2}_{j=g_1+1} u_j(p_2) \partial_j \right]^2
\Theta \left[ \begin{array}{c}  {\vec \alpha}_1 \\ {\vec \beta}_1  \end{array}
\right] (\int^z_w {\bf u} \vert \Omega_1)
\Theta \left[ \begin{array}{c}
{\vec \alpha}_2 \\ {\vec \beta}_2  \end{array} \right] (0 \vert
\Omega_2) +O(t^2)\hspace {7mm}(A.6)$$
In the above equation $v_i,u_i$ and $u_j$ are holomorphic differential
one-forms on $\sum_g , \ \ \sum_1$ and $\sum_2$, respectively and $\left[
\begin{array}{c}  {\vec \alpha}  \\{\vec \beta } \end{array} \right]=\left[
\begin{array}{ll} {\vec \alpha}_1&{\vec \alpha}_2 \\
{\vec \beta}_1&{\vec \beta}_2 \end{array} \right] .p_1(p_2)$ is the point that
i
s
generated on $\sum_1(\sum_2)$ by ZHP, $z$ and $w$ are points on $\sum_1$.
$\omega_{z-w}(p_1)$ is defined as :
$$\omega_{z-w}(x)=\partial_x ln{E(z,x)\over E(w,x)}$$
Under NZHP $\sum_{g+1}$ transforms to $\sum_g$ and Eq.(A.5) reduces
to $^{[24]}$:
\newpage
$$\Theta \left[ \begin{array}{c}  {\vec \alpha}^\prime \\ {\vec \beta}^\prime
\end{array} \right] (\int^{z_1}_{w_1} {\bf v }
\vert \Omega_{g+1}) \buildrel NZHP \over \longrightarrow
\delta_{\alpha_{g+1},0} \left(
\Theta \left[ \begin{array}{c}  {\vec \alpha} \\ {\vec \beta } \end{array}
\right]
(\int^{z_1}_{w_1} {\bf u} \vert \Omega_g)+ \right. $$
$${t^{1 \over 2} \over E(z_2,w_2)} e^{2\pi i \beta_{g+1}} \left[ R_{12} \Theta
\left[ \begin{array}{c}  {\vec \alpha} \\ {\vec \beta } \end{array} \right]
(\int^{z_1}_{w_1}
{\bf u} +\int^{z_2}_{w_2} {\bf u } \vert \Omega_g)+R_{12}^{-1} \Theta
\left[ \begin{array}{c}  {\vec \alpha} \\ {\vec \beta } \end{array} \right]
(\int^{z_1}_{w_1}
{\bf u} -\int^{z_2}_{w_2} {\bf u } \vert \Omega_g) \right] $$
$$+ \delta_{\alpha_{g+1},{1 \over 2}}
{t^{1 \over 8} \over E^{1\over 4}(z_2,w_2)} e^{\pi i \beta_{g+1}}$$
$$\hspace {1mm}\left( R_{12}^{1\over 2} \Theta \left[ \begin{array}{c}
{\vec \alpha}
\\ {\vec \beta} \end{array} \right] (\int^{z_1}_{w_1} {\bf u} + {1 \over 2}
\int^{z_2}_{w_2} {\bf u}  \vert \Omega_g)+R_{12}^{-{1\over 2}} e^{2\pi i
\beta_{
g+1}}\Theta
\left[ \begin{array}{c}  {\vec \alpha} \\ {\vec \beta } \end{array} \right]
(\int^{z_1}_{w_1}
{\bf u} -{1\over 2}\int^{z_2}_{w_2} {\bf u}  \vert \Omega_g) \right)
\hspace {7mm}(A.7) $$
In the above equation $v_i$ and $u_i$ are one-forms on $\sum_{g+1}$ and $\sum_g
$ ,$z_2$ and $w_2$ are the two new points that are created on $\sum_g$ (under
NZHP), and $\left[ \begin{array}{c}  {\vec \alpha}^\prime \\ {\vec
\beta}^\prime
\end{array} \right]=
\left[ \begin{array}{ll}  {\vec \alpha}&\alpha_{g+1} \\ {\vec
\beta}&\beta_{g+1}
\end{array} \right] $ . $R_{12}$ is defined in Eq.(45).

The other relations that we need, is the modular transformation properties of
th
e
theta functions. One can show that under a general transformation of the period
matrix, Eq.(25), $det\ \ Im \Omega $ and theta function transformed
as $^{[24]}$:
$$det \ \ Im \Omega \longrightarrow \left | det (C \Omega + D)\right |^{-2}
(det \ \ Im \Omega)$$
$$\hspace {22mm}\Theta \left[ \begin{array}{c} {\vec \alpha} \\ {\vec \beta }
\end{array}\right](0
\vert \Omega )\longrightarrow e^{-i\pi \varphi} \ \ det^{1/2} (C\Omega +D)
\Theta \left[ \begin{array}{c} {\vec \alpha}' \\ {\vec \beta }'
\end{array} \right] (0 \vert \Omega )
\hspace {22mm}(A.8)$$
where $\left[ \begin{array}{c} {\vec \alpha}^\prime \\ {\vec \beta}^\prime
\end{array} \right]$ is specified as:
$$\hspace {40mm}\left( \begin{array}{c} {\vec \alpha} \\ {\vec \beta }
\end{array}\right)= \left( \begin{array}{ll}  D&-C \\ -B&A  \end{array}\right)
\left( \begin{array}{c} {\vec \alpha}^\prime \\ {\vec \beta}^\prime
\end{array}\right)
+{1 \over 2}\left( \begin{array}{c} (CD)_d \\ (AB)_d \end{array}\right)
\hspace {25mm}(A.9)$$
In the above relation $(M)_d$ is meant a column matrix built from the diagonal
elements of $M$. In Eq.(A.8), $\varphi $ is a phase which depends on $\varphi =
\varphi (A,B,C,D,{\vec \alpha , \vec \beta })$ .
\begin{center}
{\large \sc References\\}
\end{center}
\begin{enumerate}
\item 1. A.A. Belavin, A.M. Polyakov and A.B. Zamolodchikov, Nucl. Phys. B241
(1
984)333
\item E. Witten, Commun. Math. Phys. 92 (1984)455
\item A.M. Polyakov and P.B. Wiegmann,  Phys. Lett. B141 (1984)223
\item V. Knizhnik and A.B. Zamolodchikov, Nucl. Phys. B247 (1984)83
\item D. Gepner and E. Witten, Nucl. Phys. B278 (1986)493
\item H.J. Schnitzer and K. Tsokos, Nucl. Phys. B291 (1987)429
\item S.G. Naculich and H.J. Schnitzer, Nucl. Phys. B332 (1990)583
\item E. Charpentier and K. Gawedzki, Annals of Phys. 213 (1992)233
\item M. Alimohammadi and H. Arfaei: Level-one $SU(N)$ WZNW models on HGRS,
 Int. Jour. of Mod. Phys. A (1993), to appear
\item M. Alimohammadi and H. Arfaei: $SU(N)_1$ correlation functions on
HGRS, $ibid$
\item M. Alimohammadi and H. Arfaei: Level-one $SO(N)$ WZNW models on HGRS,
Sharif Univ. Preprint, SUTDP/92/70/2
\item M. Wakimoto, Commun. Math. Phys. 101 (1986)605
\item D. Bernard and D. Felder, Commun. Math. Phys. 127 (1990)145
\item M. Frau $et al.$, Phys. Lett. B245 (1990)453 . One of the recent
papers on this subject is: H. Konno, Phys. Rev. D45 (1992)4555
\item V.G. Kac, Infinitc dimensional Lie algebras, Cambridge Univ. Press, 1985
\item See for example, P. Bouwknegt and W. Nahm, Phys. Lett. B184 (1987)359 and
the references therein
\item A.N. Schellekens and N.P. Warner, Phys. Rev. D34 (1986)3092. F.A. Bais
and P.G. Bouwknegt, Nucl. Phys. B279 (1987)561
\item P. Christe and F. Ravanini, Int. Jour. of Mod. Phys. A4 (1989)897
\item A.N. Redlich, H.J. Schnitzer and K. Tsokos, Nucl. Phys. B298 (1987)397
\item D. Mathur and A. Sen, Phys. Lett. B278 (1989)176
\item D. Gepner, Commun. Math. Phys. 141 (1991)381
\item J. Birman: Links, braids and mapping class groups, Princeton Univ. Press,
1975
\item D. Friedan and S. Shenker, Nucl. Phys, B281 (1987)509
\item See, for instance, J.D. Fay: Theta function on Riemann surfaces, Lecture
N
otes in Mathematics, Vol.352 (Springer, Berlin 1973)\\ \\
\end{enumerate}
{\bf Figure Captions} \\ \\
\begin{tabular}{ll}
Fig.1: & The fusion rule $\Phi _i \times \overline {\Phi _i} = \sum _j
  \Phi _j$ shows which fields appear as propagator of \\
 & $\Phi _i$-loop ($\overline {\Phi _i}$ stands for conjugate representation of
$\Phi _i$).\\
Fig.2: & Genus-two partition function of $\Phi_8$. In this diagram $\Phi_i$ is
$\Phi _1 , \Phi _8 , \Phi_{10}$ or $ \Phi _{\overline {10}}$.\\
Fig.3: & One of the genus-two p.fs. of $\Phi _8$.\\
Fig.4: & The cycle $a_1^{-1}a_2$ in genus-two.\\
Fig.5: & Expanding the LHS diagram in terms of the RHS diagrams (crossing
symmety).\\
Fig.6: & Genus-three partition function of $\Phi _8$. $\Phi _i$ and $\Phi _j$
ar
e $\Phi _1, \Phi _{10}$ or $ \Phi _{\overline {10}}$\\
Fig.7: & A typical diagram of $SU(3)_3$ partition function on HGRS's. $m_i$ are
$1,10$ or $\overline {10}$.\\
Fig.8: & Two-point $\Phi_8 - \Phi_8$ correlator on HGRS's.\\
Fig.9: & $2n$-point $\Pi ^n_{i=1} (\Phi _8 (z_i)- \Phi _8 (w _i))$
correlator on HGRS's\\
Fig.10: & Correlator of $<\Phi _{10} (z)>$ on the torus.\\
Fig.11: & One-point function of $\Phi _{10}(z)$ on HGRS's
\end{tabular}

\end{document}